\documentclass[9pt]{pasa}%

\usepackage{graphicx,subfig}

\title[Building extended source sky models for EoR science]{Building models for extended radio sources: implications for Epoch of Reionisation science}

\author[Trott \& Wayth]{Cathryn M. Trott$^{1,2,3}$\thanks{Email: cathryn.trott@curtin.edu.au} and Randall B. Wayth$^{1,2,3}$
\affil{$^1$International Centre for Radio Astronomy Research, Curtin University, Bentley 6845 Australia}%
\affil{$^2$ARC Centre of Excellence for All-Sky Astrophysics (CAASTRO), Curtin University, Bentley 6845 Australia}
\affil{$^3$ARC Centre of Excellence for All-Sky Astrophysics in 3 Dimensions (ASTRO 3D), Curtin University, Bentley 6845 Australia}
}%

\jid{PASA}
\doi{10.1017/pas.\the\year.xxx}
\jyear{\the\year}

\usepackage{aas_macros}
\usepackage{hyperref} 
\hypersetup{colorlinks,citecolor=blue,linkcolor=blue,urlcolor=blue}

\hypersetup{draft}

\begin{document}

\begin{frontmatter}
\maketitle

\begin{abstract}
We test the hypothesis that limitations in the sky model used to calibrate an interferometric radio telescope, where the model contains extended radio sources,
will generate bias in the Epoch of Reionisation (EoR) power spectrum.
The information contained in a calibration model about the spatial and spectral structure of an extended source is incomplete because a radio telescope cannot sample all Fourier components. Application of an incomplete sky model to calibration of EoR data will imprint residual error in the data, which propagates forward to the EoR power spectrum.
This limited information is studied in the context of current and future planned instruments and surveys at EoR frequencies, such as the Murchison Widefield Array (MWA), Giant Metrewave Radio Telescope (GMRT) and the Square Kilometre Array (SKA1-Low).
For the MWA EoR experiment, we find that both the additional short baseline $uv$-coverage of the compact EoR array, and the additional long baselines provided by TGSS and planned MWA expansions, are required to obtain sufficient information on all relevant scales. For SKA1-Low, arrays with maximum baselines of 49~km and 65~km yield comparable performance at 50~MHz and 150~MHz, while 39~km, 14~km and 4~km arrays yield degraded performance. 
%
%
%
%
\end{abstract}

\end{frontmatter}

\section{INTRODUCTION }
\label{sec:intro}
A sky model for an interferometer telescope is used for calibration, source deconvolution, and source subtraction. The sky model can be obtained: (1) externally, from prior observations with existing telescopes at the same frequency; (2) internally, via measurement with telescope itself; (3) with a combination of these. For existing experiments, the current suite of sky surveys provide the input sky models for calibration and source subtraction. These can be augmented with further surveys from the same, or upgraded versions, of existing telescopes. Complete and accurate sky models are crucial for calibrating data, and subtracting unwanted sources from the dataset. This is particularly important for Epoch of Reionisation (EoR) experiments, which aim to extract a weak signal from bright foreground contamination \citep{jacobs16,carroll16,line17,barry16}.

For the southern sky, the current suite of low-frequency sky surveys, used for building a sky model, include: the 74 MHz Very Large Array Low Frequency Sky Survey redux \citep{lane12}, the MWA Commissioning Survey \citep[MWACS,][]{hurleywalker14}, MWA GLEAM \citep{wayth15,hurleywalker17}, and GMRT TGSS \citep{intema17}.
Cross-matching tools, such as the Positional Update and Matching Algorithm \citep[PUMA,][]{line17}, combine these spatially and spectrally to provide the calibration sky model.
\citet{procopio17} explored the impact of adding GMRT TGSS information to the calibration and source subtraction model for MWA EoR in the EoR1 field, finding that the additional double-source and extended source information was important for reducing bias. An extension of that work explores the direct impact of imprecisely-modelled extended sources on the MWA EoR experiment, and we undertake that work here.

Source modelling is typically performed for physical insight into the source itself (e.g., spectral structure of radio lobes to understand their energetics) and often includes multi-wavelength information to constrain physical models. Such studies typically rely on image-plane maps \citep[e.g.,][and references therein]{perley84,salter89,castelletti07,braun13,mckinley15,procopio17}. \citet{perley84}, for example, studies the multi-frequency spatial structure of Cygnus A.
The image data are represented on a discretized grid at a set of frequencies, and this forms a representation of the underlying information, some of which has been lost. This is a commensurability problem, whereby continuous data are represented discretely.
\citet{braun13}, for example, then uses the spatial power in these frequency slices to estimate the structural properties of the source. A subset of the full data has therefore been used for the analysis.

Image reconstruction and analysis tools are designed with minimal information loss considerations, but these are imperfect. Performing analysis in the measurement plane offers the greatest ability to preserve information, but can be computationally and algorithmically challenging compared with the image plane. In this work, we consider the \textit{full} information content of the data, taking this as the best possible outcome (most optimistic), while appreciating that using a segmented image-plane representation will likely degrade the result. In Section \ref{sec:methods} we explore the impact of discretization more formally.

The SKA aims to be the world's foremost radio telescope, and aims to make major scientific advances across a range of programs. The EoR and Cosmic Dawn experiment \citep{koopmans15} is envisaged to be one of the most challenging, and is one of the SKA High Priority Science Objectives. Prior to its construction and commissioning, existing and past facilities are providing the sky models for the current generation of southern hemisphere EoR experiments, such as the MWA \citep{tingay13_mwasystem}, the Precision Array for Probing the Epoch of Reionization (PAPER){\footnote[1]{http://eor.berkeley.edu}} \citep{parsons10}, and Hydrogen Epoch of Reionization Array (HERA){\footnote[2]{http://reionization.org}} \citep{deboer17}.
Compared with the SKA, current low frequency sky surveys have poorer sensitivity and spatial resolution, limiting the ability of these surveys to accurately measure the spatial and spectral structure of complex sources.

For SKA1-Low, with its long baselines and good snapshot $uv$-coverage, we expect that SKA1-Low itself will form the primary sky model.
Therefore, the sky model will be well-sampled on modes measured by SKA, and not well-measured on scales that are not.
We note that, in general, the application to calibration of an incomplete sky model formed exclusively by the same instrument being calibrated, will likely produce `re-substitution bias', where the performance of a calibration procedure is over-estimated (similar to the re-use of a training set for the real dataset in machine learning).

%
%

In this paper, we discuss the impact of the current and future suite of sky surveys at EoR frequencies to enable EoR and Cosmic Dawn (CD) science, with particular reference to the EoR and CD power spectrum. In the current era, we study the availability of sufficient information from existing and imminent sky surveys to measure the values for the parameters describing the spatial and spectral structure of extended sources.
%
%
For the future, we discuss the implications for the planned SKA1-Low array configuration to execute its ambitious EoR/CD program. We use data information content, as the basis for quantitatively assessing the performance of the instrument under a set of defined designs.
We begin with the original MWA Phase I design, and subsequently add TGSS, and the additional baselines of the upgraded MWA, and show whether the EoR science experiment is biased by the incomplete information available to correctly represent extended sources. 
Attention is then turned to the future SKA1-Low array, and the more challenging EoR/CD experiments proposed. We use these results to inform the smallest array maximum baseline required to execute these experiments.

\section{Methodology}\label{sec:methods}
Calibration of a radio interferometer requires estimation of the unknown complex gain parameters of each station. It typically relies on the fitting of data to a model of the sky signal plus instrument, allowing freedom in the complex gain parameters to perform the least-squares fit. Successful calibration therefore requires good knowledge of the received sky signal and the instrument (station locations, sky response, etc.). This is partially true for purely redundant arrays (e.g., PAPER, HERA), where sky information is used for initial calibration estimates and breaking degeneracies \citep[e.g., Omnical;][]{ali15}

In this work, the Fisher Information is used to quantify the ability of a given array configuration to estimate the sky model parameters for a generalised extended source, in the presence of a realistic background of point sources.
We use the residual model uncertainties as a measure of the error (uncertainty) in the model, and propagate these uncertainties into visibilities measured by the telescope for EoR science. We further propagate these uncertainties into the EoR power spectrum, yielding a measure of the error (power bias) due to the incomplete source model. In studying different SKA array configurations, we focus on removal of outer stations to reduce the maximum baseline, but do not re-locate stations to the array core. For the generalised extended source, we are careful to model structure on all scales of relevance for the SKA, in order to robustly and generally assess calibration performance (scales from the size of the PSF to the FOV). We apply the same model throughout, for the SKA and also the MWA and TGSS-derived results. This is to ensure consistency, and to test the ability of precursor instruments to form a sky model of relevance for the future SKA. For the MWA and TGSS, the existence of power on scales much finer than those available to their baselines, will test whether unmeasured power on small scales impacts the estimation of longer wavemodes, of relevance to the EoR.

\subsection{Approach}
We use the Fisher Information, and then the Cramer-Rao Bound \citep[CRB]{kay93} to quantify the information available in our calibration data to estimate the values of the spectral and spatial parameters describing a generalised extended source, embedded within measurement noise, and a realistic sky of extragalactic point sources.
The calibration model is obtained from the sky survey used to construct the sky model.

The Fisher Information computes the amount of information a given dataset (with a particular probability distribution function of noise, here modelled to include radiometric and background sources) contains about the values of a model parameter, for a pre-defined source model.
In general, measurements that vary rapidly with a varying parameter value have the ability to estimate its value precisely.
In contrast, no change of the expected measurement with a varying parameter value means that there is no information available to estimate that parameter. The CRB takes into account degeneracies between model parameters and correlations in the data, and represents the estimation performance of an ideal estimation algorithm.

For complex-valued data, embedded within generalised Gaussian noise with covariance, $\boldsymbol{C}$, and an expected signal vector $\vec{\mu}=\vec{\mu}(\vec{\theta})$ with parameters $\vec{\theta}$, the Fisher Information Matrix has elements:
\begin{equation}
[I]_{ab} = \left( \frac{\partial\vec{\mu}}{\partial{\theta}_a} \right)^\dagger \, \boldsymbol{C}^{-1} \, \left( \frac{\partial\vec{\mu}}{\partial{\theta}_b} \right),
\label{eqn:fim}
\end{equation}
where $ab$ are two elements of the parameter vector. In general, the data covariance can also be a function of unknown parameters, but here we assume we have full knowledge of the per-visibility properties of the radiometric noise:
\begin{equation}
{_{th}{\bf C}}(u,v;\nu,\nu^\prime) = \left(\frac{2k{\rm T_{sys}}}{{\rm A_{eff}}}\right)^2\frac{1}{{\Delta\nu\Delta{t}}} \delta(\nu-\nu^\prime) \,\,{\rm Jy}^2,
\end{equation}
and the point source covariance \citep{trott16} as a function of spectral channels $\nu, \nu^\prime$ and Fourier mode $\vec{u}=u,v$:
\begin{eqnarray}
_{fg}{\bf C}(\vec{u};\nu,\nu^\prime) &=& \frac{\alpha}{3-\beta}\frac{S_{\rm max}^{3-\beta}}{S_0^{-\beta}}\frac{\pi{c^2}\epsilon^2}{D^2} \frac{1}{\nu^2 + \nu^{\prime{2}}}\nonumber\\
&\times& \exp{\left( \frac{-|\vec{u}|^2c^2f(\nu)^2\epsilon^2}{4(\nu^2 + \nu^{\prime{2}})D^2} \right)} \, {\rm Jy}^2,
\end{eqnarray}
where $\epsilon=0.42$ converts an Airy disk to a Gaussian characteristic width, $D$ is the station diameter, and $f(\nu) = (\nu-\nu^\prime)/\nu_0$.
The point source model is represented by a broken power-law, characterised by parameters $\alpha,\beta$, such that in a sky area $d{\bf l}$,:
\begin{eqnarray}
\langle{N(S,S+dS)}\rangle(\nu) &=& \frac{dN}{dS}(\nu)\,dS\,d{\bf l} \\&=& \alpha \left( \frac{\nu}{\nu_0} \right)^{\gamma} \left( \frac{S_{\rm Jy}}{S_0}\right)^{-\beta}\,dS\,d{\bf l}.
\label{source_counts}
\end{eqnarray}
We use values of $\alpha=4100\,{\rm Jy}^{-1}{\rm sr}^{-1}$, $\beta=1.59$ and $\gamma=-0.8$ at 150~MHz \citep{intema11}.
We assume that our sky model is formed from 15~minutes (1~hour) of data for GMRT (MWA), reducing the radiometric noise component with respect to the confusion, and for spectral resolution commensurate with the EoR experiment (100~kHz). (The GMRT TGSS survey used 15-min per field, split over 3--5 pointings.)
Observations are assumed to be of a field at declination -27 degrees, centered on HA=0.

The CRB yields the uncertainties, and correlations, for each parameter value. If we assume that these uncertainties are then embedded within the sky model for that extended source, we can propagate these errors into the science data measured by the telescope (the measured visibility dataset). This is achieved using a standard Jacobian and the Fisher Information, $I$, such that:
\begin{equation}
{\bf C}_V(u,v;\nu) = \boldsymbol{J}^\dagger \, \boldsymbol{I}^{-1} \, \boldsymbol{J},
\label{eqn:covvis}
\end{equation}
where $\boldsymbol{J}$ is the matrix of derivatives of the visibility measured at $u,v$ and channel $\nu$. We further propagate from measured visibilities to the EoR power spectrum;
\begin{equation}
\Delta{P}(k_\bot,k_\parallel) = \left(  \mathcal{F}_\nu^\dagger \, \mathcal{W}^\dagger {\bf C}_V \mathcal{W}\, \mathcal{F}_\nu \right) \, \delta(k_\parallel-k_\parallel^\prime,k_\bot^2-u^2-v^2),
\end{equation}
where $\mathcal{F}_\nu$ is the Fourier Transform operator along the spectral direction, $\mathcal{W}$ is a spectral taper function that aims to reduce spectral leakage, and the delta-function extracts the variance estimates (power) from the covariant matrix as well as identifying the perpendicular $k$ modes with the $L_2$-norm of the angular Fourier modes ($k_\bot^2=u^2+v^2$). Herein we employ a Blackman taper.

Therefore, we can assess the error introduced into our science data by an incomplete extended source model produced from a given array configuration.

Fourier Transform of the data to the image plane, and discretization into surface brightness pixels, transforms the visibility covariance matrix (equation \ref{eqn:covvis}) into a degraded image covariance, such that:
\begin{equation}
{\bf C}_I(l_i,m_i;\nu) = \boldsymbol{\mathcal{D}_i}^\dagger \boldsymbol{\mathcal{F}}^\dagger \boldsymbol{J}^\dagger \, \boldsymbol{I}^{-1} \, \boldsymbol{J} \boldsymbol{\mathcal{F}} \boldsymbol{\mathcal{D}_i},
\end{equation}
where $\boldsymbol{\mathcal{D}_i}$ and $\boldsymbol{\mathcal{F}}$ denote the discretization operator (continuous-to-discrete) and spatial Fourier operator (discrete-to-continuous), respectively, and the Jacobians act on the data used to estimate the source properties. Inadequate discretization sampling compared with scales and shapes of the underlying source components, will degrade the quality of the estimation, and this can be quantified by studying the mean-squared-error (MSE) between the actual underlying source structure and the discretized representation. This can be seen most simply by considering a Fourier Transform back to the visibility plane (to estimate scale sizes), where a discrete-to-discrete transform couples pixel properties into the data covariance matrix. At this point, both information loss and bias in the estimates, are possible.

\subsection{Generalised extended source model}
We want to form the most general extended source, in order to provide a fair basis for comparison. We aim to build a model with multi-scale structure that encases the spatial modes accessible to the Baseline Design SKA1-Low, corresponding to scales of the synthesized beam (10~arcsec at 150~MHz) to a fraction of the field-of-view (0.08~degrees). We build a model with a complete set of angular scales, spaced evenly in the range $k=$~[570--17000]~rad$^{-1}$, and construct a source based on the summation over a series of Gaussians, each with five parameters: central location $l_i,m_i$ (rad), peak brightness $S_i$ (Jy/beam for SKA-Low beam at 150~MHz), characteristic scale $\sigma_i$ (rad), and spectral index $\gamma_i$. This corresponds to 57 angular scales ($k=$~[570--17000]~rad$^{-1}$, sampling evenly at a spacing of half of the lowest $k$-mode), yielding 285 total parameters. The parameters for each scale are generated via a Gaussian-distributed random sampling, as described in Table \ref{table:bubbles}, and fixed thereafter for each array configuration.
\begin{table}
\centering
\begin{tabular}{|c||c|}
\hline 
Parameter & Value (scale $i$) \\ 
\hline \hline
$a_i$ (Jy/bm) & $1/\sqrt{1+i}$ \\ 
\hline 
$l_i$ (arcmin) & $\mathcal{N}$(0,$\Delta{l}^2=10^2$)\\ 
\hline
$m_i$ (arcmin) & $\mathcal{N}$(0,$\Delta{m}^2=10^2$)\\ 
\hline 
$\sigma_i$ (rad) & 1/$k_i$ \\ 
\hline 
$\gamma_i$ & $\mathcal{N}$($-$0.8,$\Delta{\gamma}^2=0.02^2$) \\ 
\hline  
\end{tabular}
\caption{Parameter values as a function of scale, $k_i$. $\mathcal{N}(\mu,{\rm var})$ denotes a Gaussian-distributed random number with mean $\mu$ and standard deviation $\sqrt{{\rm var}}$.}\label{table:bubbles}
\end{table} 
The Fourier-space expectation of the signal ($\mu$ in Equation \ref{eqn:fim}) is given by:
\begin{eqnarray}
\mu(\vec{u},\nu;\vec{\theta}) &=& \sqrt{2\pi} \displaystyle\sum_{i=1}^{57} \, a_i \sigma_i^2 \left(\frac{\nu}{\nu_0}\right)^{\gamma_i} \nonumber\\ 
&\times& \exp{(-2\pi{i}\vec{u}\cdot\vec{l}_i)} \, \exp{(-2|u|^2\pi^2\sigma_i^2)},
\end{eqnarray}
where the two exponentials encode the Fourier kernel, and the multi-scale Gaussian, respectively, and $\vec{\theta}=[a,l,m,\sigma,\gamma]$.

\begin{figure}
\includegraphics[width=.45\textwidth]{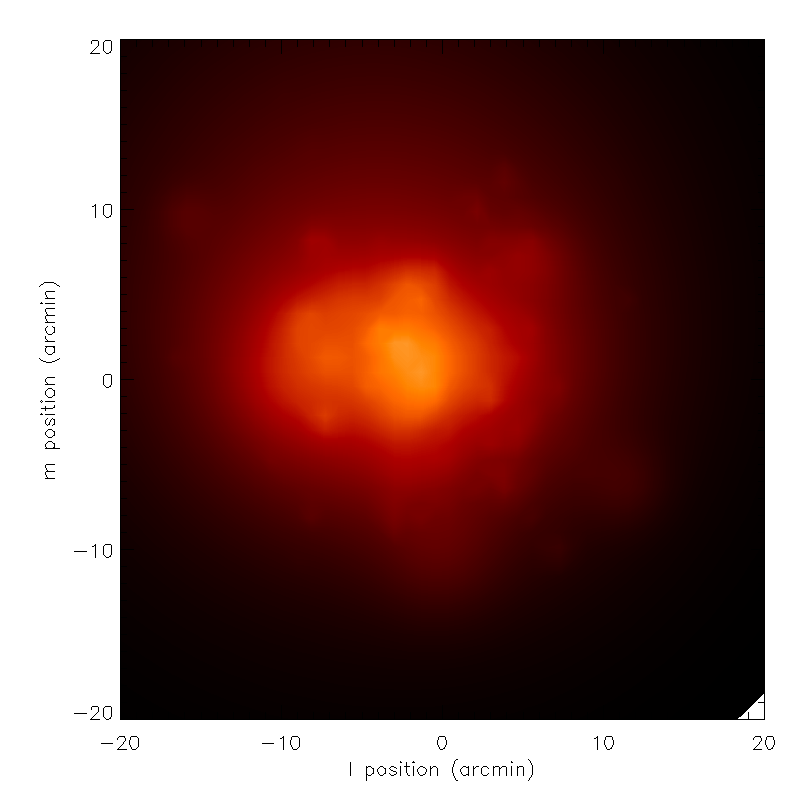}
\caption{Image of the extended source produced using the model described (lowest spectral channel).}
\label{fig:image}
\end{figure}

Figure \ref{fig:image} shows an image of the complete model.
Note that, \textit{although a base source peak brightness of 1~Jy/beam has been chosen for the source, the amplitude here is not relevant to the final error introduced into the visibilities. This is because brighter sources can be estimated more precisely, and weaker sources less precisely, with a linear scaling with peak brightness. The propagation of error back into the visibilities also scales linearly with peak brightness, and these scalings cancel \citep{trott12}. Nonetheless, the peak brightness refers to an SKA-Low beam at 150~MHz, and remains consistent for all models.} Therefore, each extended source with these angular scales being estimated for the sky model will contribute this error.

\subsection{Extended sources within the field}

The procedure described above yields the power bias from a single extended source in the sky model for a given field. If we now consider the total number of extended sources in a given field, we can estimate the total power bias in the EoR power spectrum to be $N_{\rm src}\Delta{P}(k_\bot,k_\parallel)$, where $N_{\rm src}$ is the number in the field. This implies that the extended sources considered here all have the same power spectrum of spatial and spectral structure, although each individual source will have a different realisation of the parameter values. This ensures we are considering a consistent set of sources.

There is no complete low-frequency census of extended sources, due to the limited sensitivity and spatial resolution under study in this work. Results from the high-resolution FIRST \citep[1.4~GHz)][]{becker94} and TGSS catalogues \citep{intema17} can be used to estimate the number of sources per unit area of a given angular scale and flux density. \citet{procopio17} showed that the brightest sources were most important to model, because these tended to be closer and therefore more extended.

\begin{figure}
\includegraphics[width=.45\textwidth]{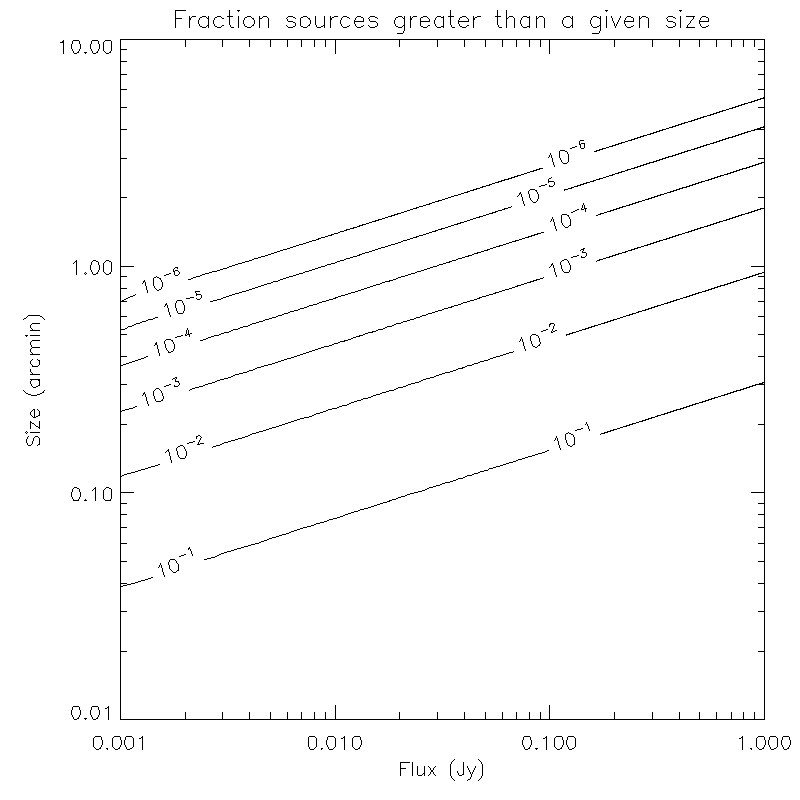}
\caption{Fraction of sources with angular extent greater than a given size, as a function of source flux density, for 150~MHz \citep[adapted from][]{windhorst90}.}
\label{fig:fraction}
\end{figure}

\begin{figure*}
\subfloat{
\includegraphics[width=.32\textwidth]{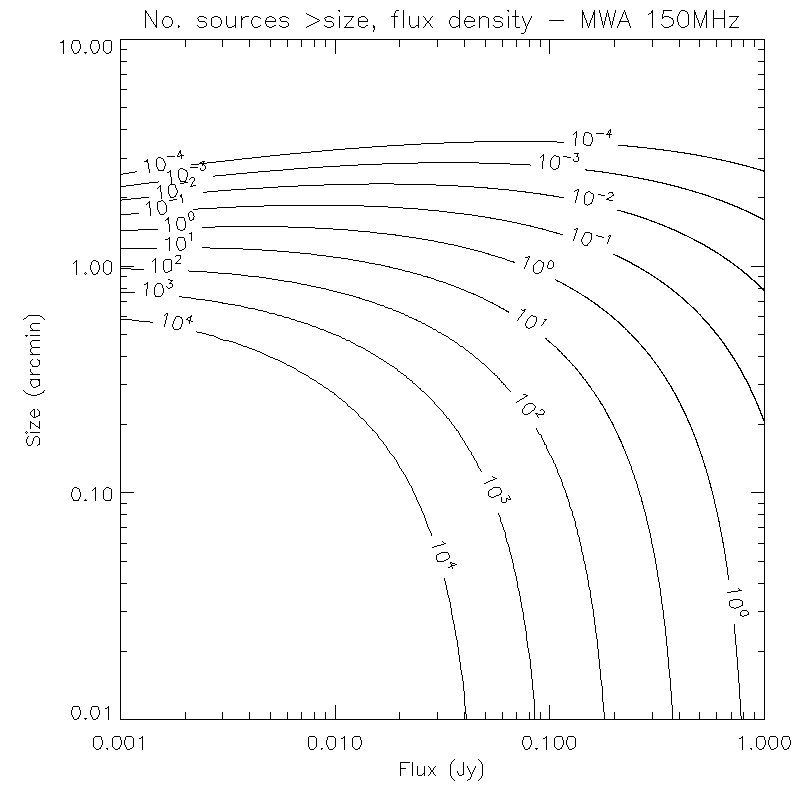}
}
\subfloat{
\includegraphics[width=.32\textwidth]{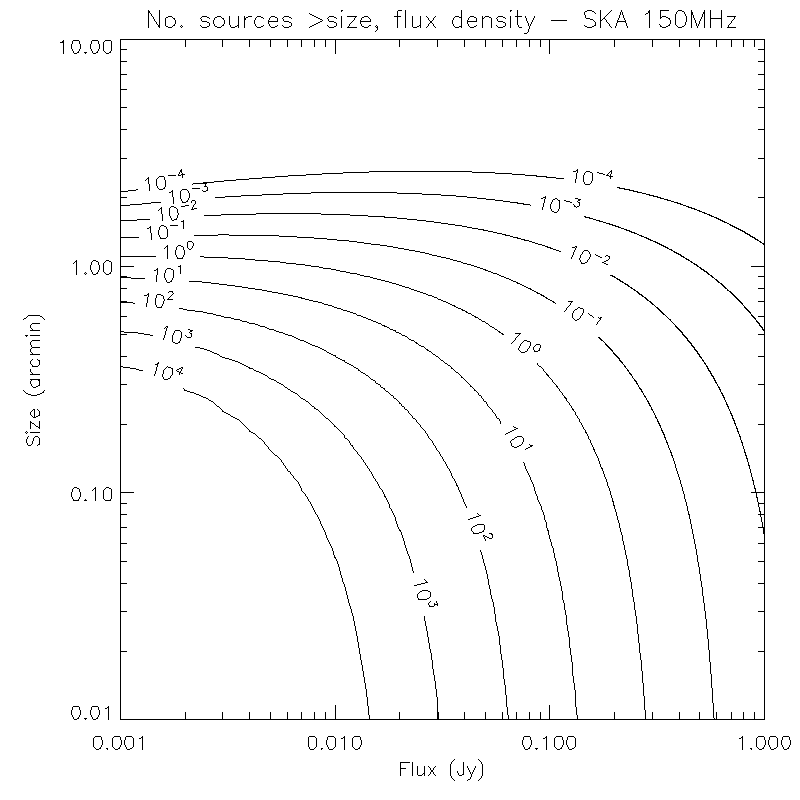}
}
\subfloat{
\includegraphics[width=.32\textwidth]{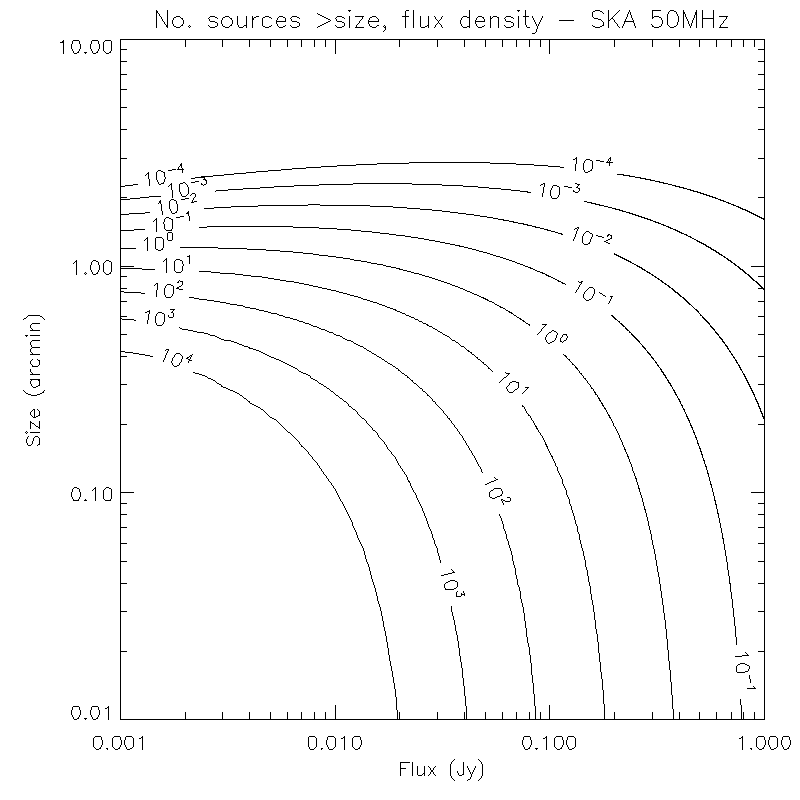}
}
\caption{Number of sources with angular extent and flux density greater than given values for the MWA 150~MHz (left), SKA 150~MHz (centre), and SKA 50~MHz (right) experiments.}
\label{fig:number}
\end{figure*}

We use the study of \citet{windhorst90} at 1.4~GHz to estimate the 150~MHz and 50~MHz distributions, assuming a spectral index of $-0.8$ between the bands and no structural evolution. They find that the fraction of sources larger than size $\psi$ (arcsec) at flux density, $S$ (mJy), can be described by:
\begin{equation}
h(\psi) = \exp{[-\ln{2}(\psi/\psi_m)^{0.62}]},
\end{equation}
where $\psi_m = 2.0\,S_{1.4}^{0.3}$~arcsec is the median size at 1.4~GHz. We scale the flux densities to 150~MHz to find the corresponding median size for low frequencies. Figure \ref{fig:fraction} shows this fraction as a function of flux density and size. Strictly this distribution is applicable for flux densities below 1~Jy: the extended, close radio galaxies (e.g., Fornax A, Centaurus A, Pictor A), which have much larger flux densities, have their own individual spatial and spectral distributions.
Assuming a point source number count distribution, a field-of-view and this fraction, the number of sources larger than a given angular size and greater than some flux density limit can be approximated as:
\begin{equation}
N(>\psi,S_0) = \displaystyle\int_{S_0}^\infty \frac{dN}{dS}\,dS\,\Omega \, h(\psi) = \displaystyle\int_{S_0}^\infty \alpha S^{-\beta}\, \Omega \, h(\psi),
\end{equation}
where $\alpha = 4000 (\nu/150)^{-0.8}$~Jy$^{-1}$sr$^{-1}$, $\beta=\{1.59~(S<1~{\rm Jy}),~2.5~(S>1~{\rm Jy})\}$, parametrise the number count distribution \citep{intema11}, and $\Omega$ is the field-of-view (steradians). Figure \ref{fig:number} shows the associated contour plots for the MWA 150~MHz, SKA 150~MHz, and SKA 50~MHz experiments.
At the flux density confusion limits of these experiments, we would expect $\sim$1 source in each of the experiments, of scale larger than $\sim$1~arcmin. Most of the weaker sources have angular extents of 10s of arcseconds, and these have fewer critical parameters for estimation. Given these estimates, we focus on the bright, distributed nearby radio galaxies in this work, and consider a single source of importance in the field.
%
%


\subsection{Telescope arrays}
\subsubsection{Murchison Widefield Array}
\begin{figure*}
\subfloat{
\includegraphics[width=.48\textwidth]{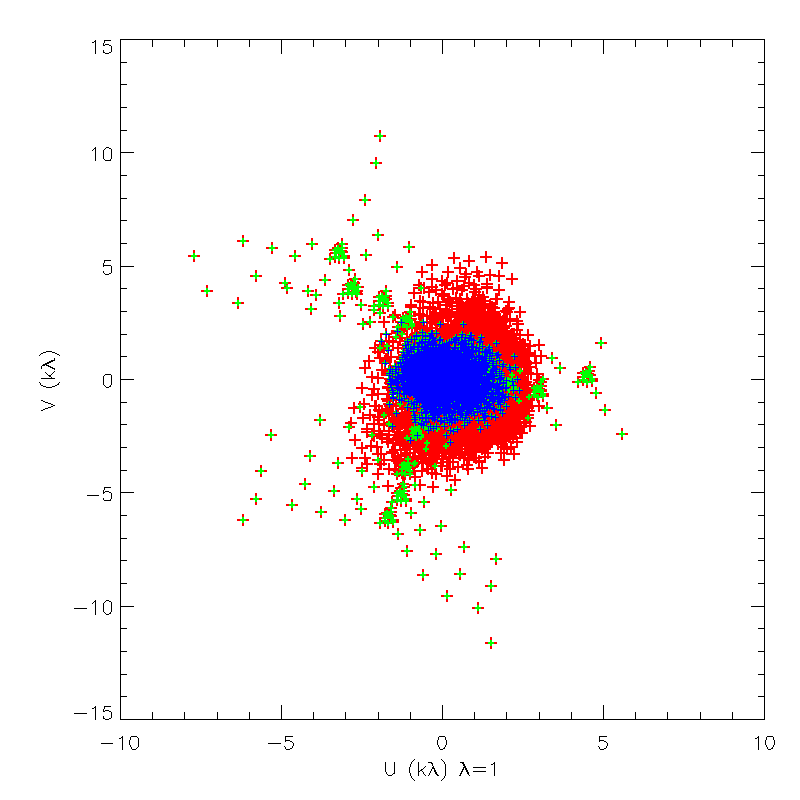}
}
\subfloat{
\includegraphics[width=.48\textwidth]{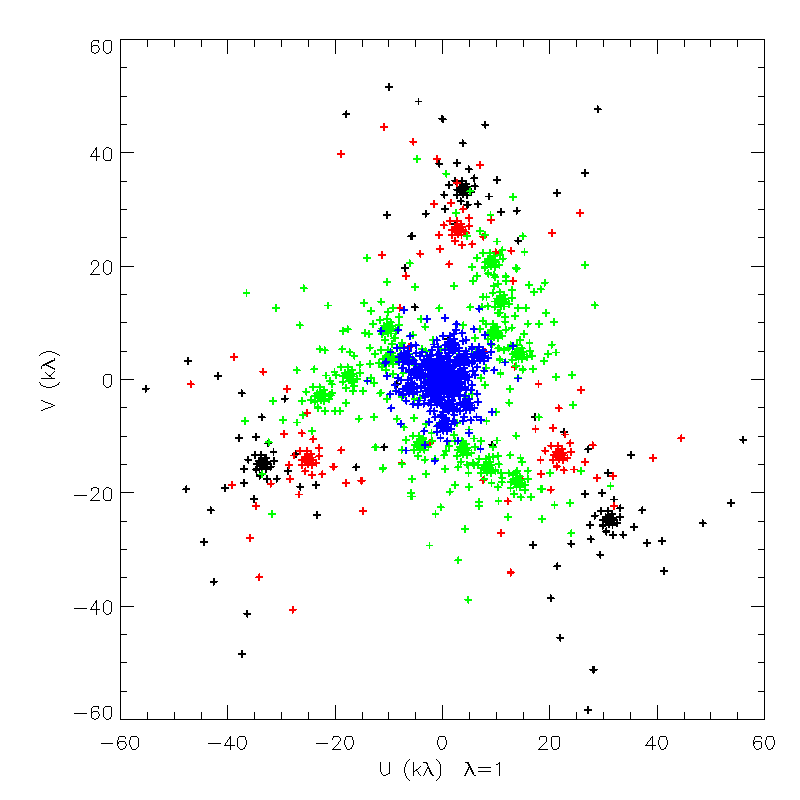}
}
\caption{(Left) $uv$ coverage for a zenith snapshot pointing for the three MWA arrays: MWA Phase I (blue), +GMRT (green), MWA Phase III+GMRT (red). (Right) $uv$ coverage for a zenith snapshot pointing for the four SKA1 arrays considered: black (max. baseline 65~km), red (49~km), green (39~km), blue (14~km), ($\lambda=1$~m).}
\label{fig:uv}
\end{figure*}

The first three years of MWA EoR observations form the basis for all published limits to date, and use the Phase I MWA configuration with 128 tiles. The GLEAM survey uses the same array for its observations. The GMRT TGSS survey is a re-processing of the GMRT 150~MHz sky survey with new methods for instrument and ionospheric calibration \citep{intema17}. The additional sensitivity of the GMRT, and the longer baselines, provide higher spatial resolution, but the small number of dishes (30) limits the additional $uv$-coverage. The upgraded MWA will provide additional surface brightness sensitivity and spatial resolution with its 256 tiles. The zenith-pointed $uv$-coverage for these are displayed in Figure \ref{fig:uv} (left). GMRT clearly adds information at high spatial resolution, but even with rotation synthesis, the coverage is sparse and the larger angular scales are not significantly improved.

GMRT has different system temperature, latitude and dish sensitivity (effective area) compared with the MWA, and these are taken into account in the analysis. Earth Rotation synthesis is considered when computing the $uv$-coverage, assuming an EoR field-of-interest located at declination $-$27~degrees (MWA latitude). The GMRT TGSS survey parameters (15 minutes integration time per field) are used as the basis for the sky model formed from these different hybrid arrays.


\subsubsection{SKA1-Low configurations}
We trial four array configurations, which are all subsets of the Baseline Distribution. Table \ref{table:array} describes the number of stations, and maximum baseline, for each.
\begin{table}
\centering
\begin{tabular}{|c||c|}
\hline 
$N_{\rm stn}$ & Max. baseline (km) \\ 
\hline \hline
512 & 65 \\ 
\hline 
494 & 49 \\ 
\hline
476 & 39 \\ 
\hline 
400 & 14 \\ 
\hline
\end{tabular}
\caption{The four arrays considered.}\label{table:array}
\end{table} 
Figure \ref{fig:uv} (right) shows the zenith snapshot uv-coverage of each at 150~MHz.
We also considered a 4~km baseline array, which effectively corresponds to an array with EoR-science scales. Additionally, we consider a variation on the Baseline Design, whereby the inner clusters of six stations are unpacked for improved instantaneous imaging performance \citep{jones16}.

The station effective areas, and sky temperature as a function of frequency, are taken from the SKA1-Low system description (L0 Requirements). We compute the EoR/CD power spectrum error at two prime frequencies of interest (150~MHz, $z$=8.6; 50~MHz, $z$=27), and compare to typical expected 21~cm cosmological power spectra, from 21cmFAST simulations \citep{mesinger11}. Standard conversions are undertaken from Jy$^2$~Hz$^2$ to mK$^2h^{-3}$~Mpc$^3$ \citep{morales04}.

\section{Results}\label{sec:results}
We form the power biases due to extended sources requiring estimation for formation of the sky model, and compare these with expected 21~cm power spectra obtained from 21cmFASTv2 (reionisation via faint galaxies).

\subsection{MWA}
\begin{center}
\begin{figure*}
\includegraphics[width=.95\textwidth]{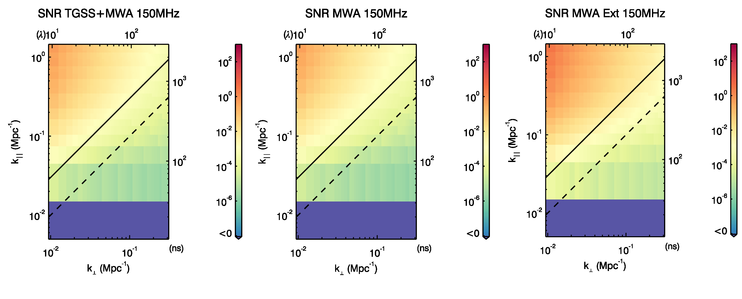}
\caption{Signal-to-noise (contrast) ratios of a typical 21~cm cosmological signal to the power error introduced by one extended source in the field, for the three MWA-based hybrid array configurations considered, and a 10~MHz bandwidth experiment centred at 150~MHz ($z=8.6$).}
\label{fig:mwa}
\end{figure*}
\end{center}

The MWA EoR experiment aims to detect the 21~cm signal through a power spectrum at redshifts, $z=6.5-9.0$ ($\nu=137-197$~MHz). We take a nominal lower frequency of 150~MHz and estimate the power bias due to the three hybrid arrays. Figure \ref{fig:mwa} displays the signal-to-power bias ratio (SNR).

Addition of the small scales from TGSS does not provide a substantial improvement in performance on EoR scales, and this hybrid array and the original MWA 128-tile array display contrast ratios of order 0.1 in the lowest portion of the EoR window. Extending the array to 256 tiles and including TGSS increases the $uv$-coverage on all scales, leading to an improvement and contrast ratios exceeding unity across the EoR window.

\begin{figure*}
\vspace{-0.5cm}
\subfloat{
\includegraphics[width=.44\textwidth]{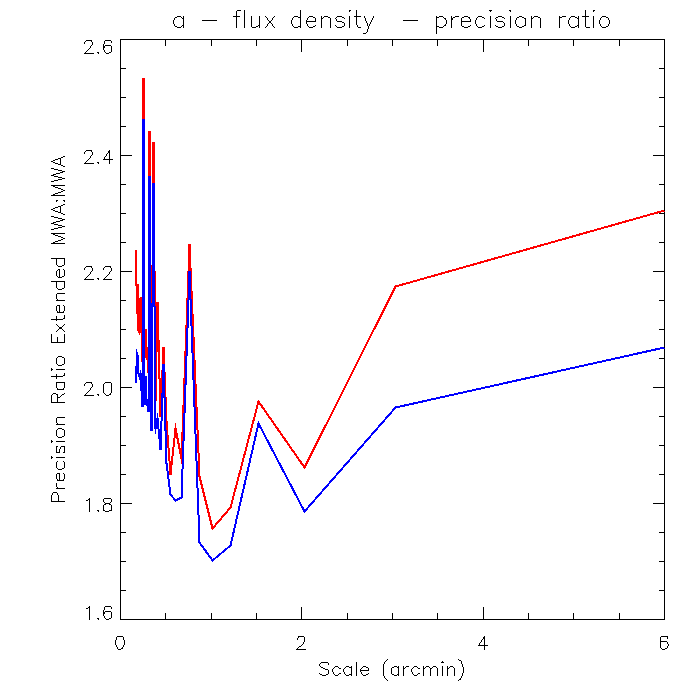}
}
\subfloat{
\includegraphics[width=.44\textwidth]{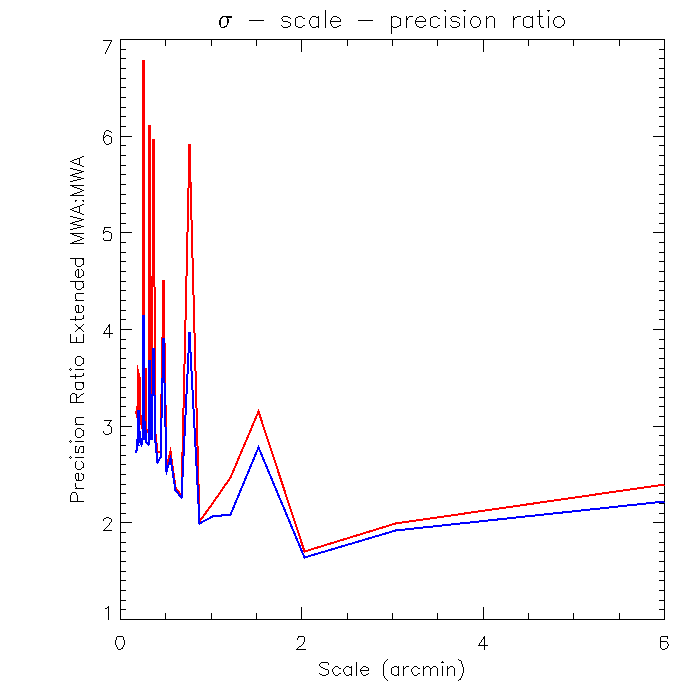}
}\vspace{-0.5cm}\\\vspace{-0.5cm}
\subfloat{
\includegraphics[width=.44\textwidth]{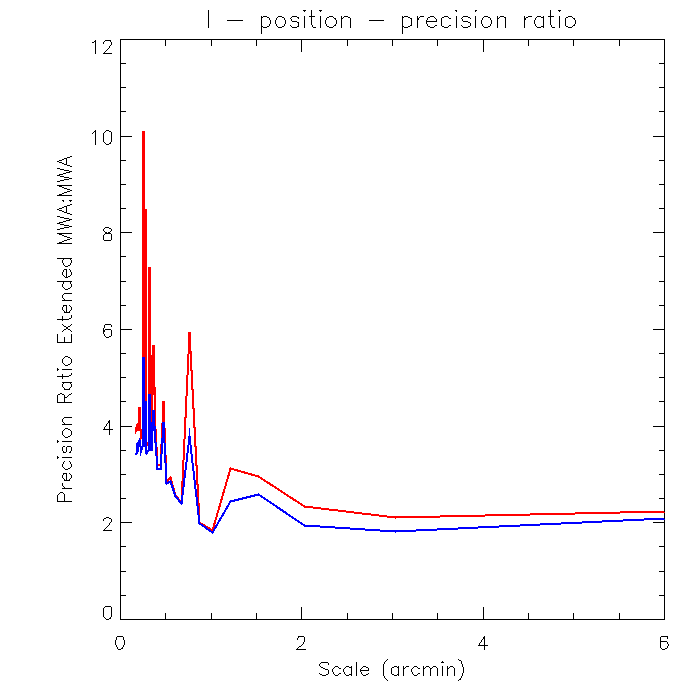}
}
\subfloat{
\includegraphics[width=.44\textwidth]{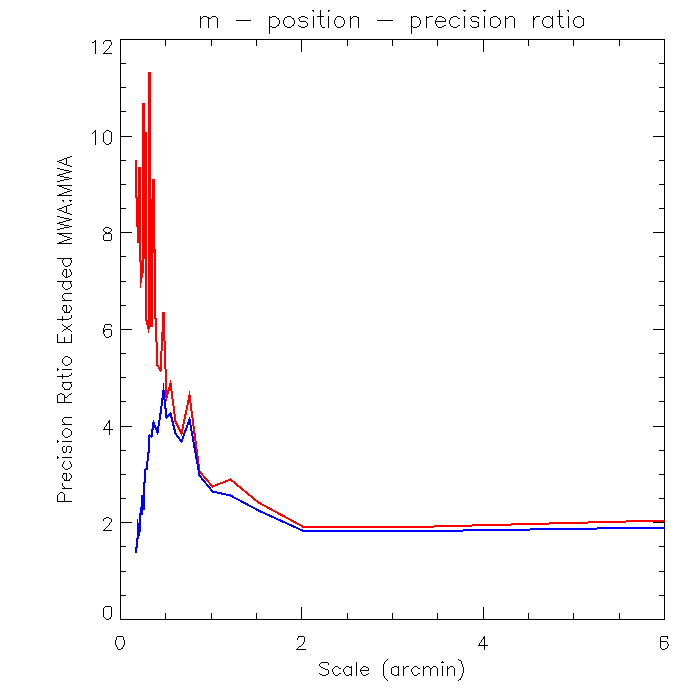}
}\\\vspace{-0.25cm}
\subfloat{
\includegraphics[width=.44\textwidth]{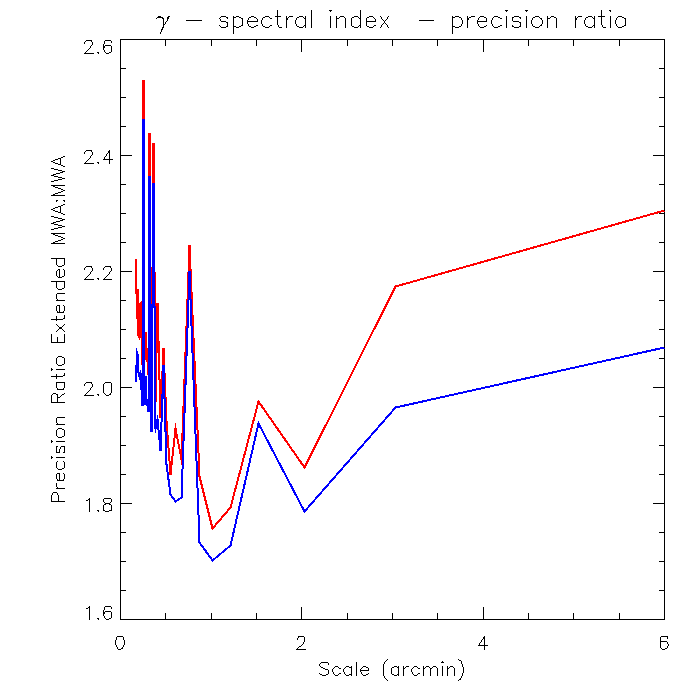}
}
\subfloat{
\includegraphics[width=.44\textwidth]{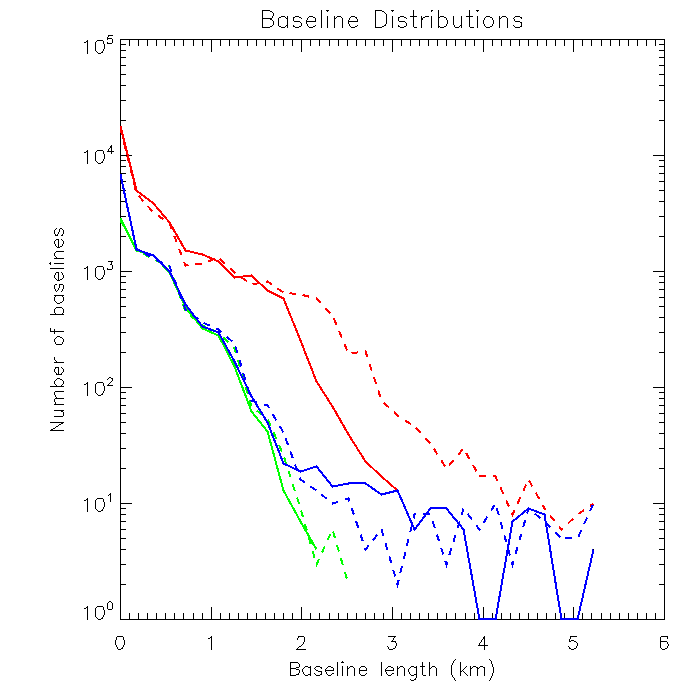}
}
\caption{(Top, middle, bottom left) Ratio of estimation performance (precision) for extended MWA (256 tiles + TGSS) relative to the original MWA128 (red) and MWA128+TGSS (blue), as a function of scale of source feature. (Bottom right) Histograms of baseline distributions in $u$ (solid) and $v$ (dashed) directions for MWA256+TGSS (red), MWA128+TGSS (blue) and MWA128 (green).}
\label{fig:param_ratios}
\end{figure*}

The improvement in parameter estimation can be explored to understand the relative importance of $uv$-coverage and maximum baseline in measuring the source parameters. Figure \ref{fig:param_ratios} display results for the five source parameters as a function of angular scale of the feature ($\sigma$). Each plot shows the ratio of estimation performance (square-root of CRB) for the extended MWA (256 tiles + TGSS) relative to the original MWA128 and MWA128+TGSS.
Also displayed are histograms of the $u$ (solid) and $v$ (dashed) distributions for each array.

It is clear that the additional long north-south baselines for the MWA256+TGSS offer improved performance on small scales for estimating the $m$-position of each source feature. It is also clear that the increased number of short baselines from the MWA256 hexagonal arrays improves estimation of large scale features. Addition of the TGSS long baselines aids the precision. The clear and intuitive conclusion is that for sources with information on a range of angular scales, both long baselines and good short baseline coverage are required in both dimensions. Particularly for EoR scales, excellent short baseline $uv$-coverage is crucial.

\subsection{SKA1-Low}
Figures \ref{fig:150mhz_result} and \ref{fig:50mhz_result} display the signal-to-noise ratios (power contrast ratios) for the four arrays considered. The dashed and solid black lines denote the first sidelobe and horizon limits for the expected leakage of foreground power (the `wedge').
\begin{center}
\begin{figure*}
\includegraphics[width=.85\textwidth]{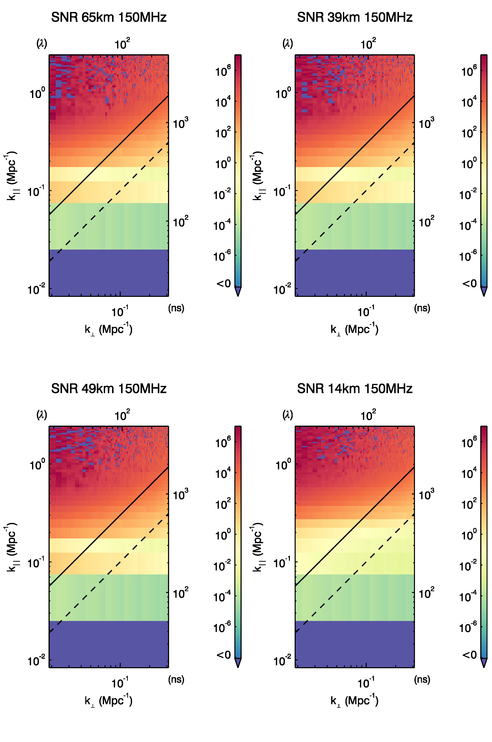}
\caption{Signal-to-noise (contrast) ratios of a typical 21~cm cosmological signal to the power error introduced by one extended source in the field, for the four array configurations considered, and a 10~MHz bandwidth experiment centred at 150~MHz ($z=8.6$).}
\label{fig:150mhz_result}
\end{figure*}
\begin{figure*}
\includegraphics[width=.85\textwidth]{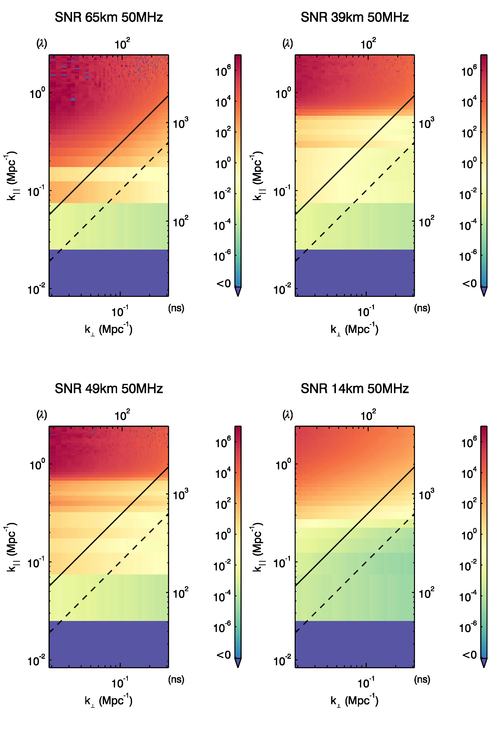}
\caption{Signal-to-noise (contrast) ratios of a typical 21~cm cosmological signal to the power error introduced by one extended source in the field, for the four array configurations considered, and a 10~MHz bandwidth experiment centred at 50~MHz ($z=27$).}
\label{fig:50mhz_result}
\end{figure*}
\end{center}
At 150~MHz, all four arrays yield SNR$>$10 across most of the EoR Window (area outside of the horizon wedge). The Blackman taper performs well to suppress foreground leakage, but at the expense of a broader DC term, and residual error leaks into the EoR window at $k_\bot\simeq{0.02}, k_\parallel\simeq{0.1}$~Mpc$^{-1}$ for all arrays. The relative strength of the 21~cm signal to the sky temperature yields acceptable performance for the 49~km and 65~km arrays, while degradation is evident for 39~km and 14~km. \textit{Notably, the 49~km and 65~km arrays yield comparable results (power ratio $\simeq$~1.004)}.

At 50~MHz, the system temperature is higher, foregrounds are brighter, and the signal is weaker. Therefore, the performance is degraded for all arrays, relative to 150~MHz. Both the 14~km and 39~km arrays yield a low SNR detection across a large region of the EoR window, while 49~km and 65~km yield good performance (SNR$>$10$^3$). Again, the 49~km and 65~km arrays yield acceptable performance (high contrast ratios).

\begin{figure*}
\subfloat{
\includegraphics[width=.48\textwidth]{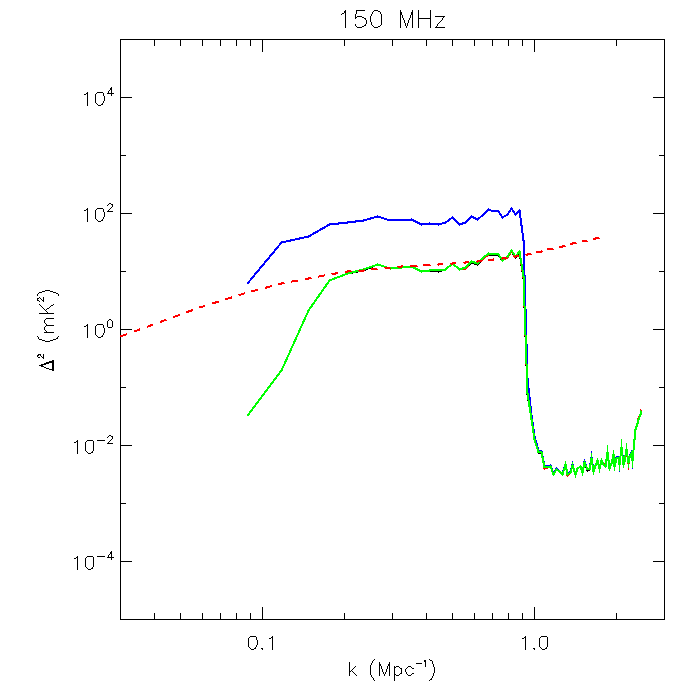}
}
\subfloat{
\includegraphics[width=.48\textwidth]{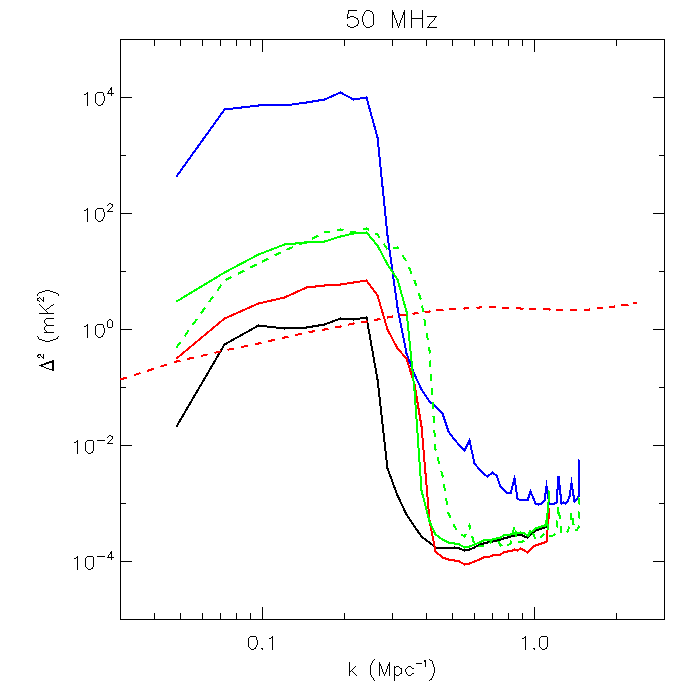}
}
\caption{(Left) 150~MHz spherically-averaged power spectrum and simulated 21cmFAST 21~cm power for comparison (dashed). 14~km (blue), 39~km (green), 49~km (red), 65~km (black). Note that the black, red and green are overlapping. (Right) Same but for 50~MHz, including an alternative array configuration with 39~km baselines (green, dashed).}
\label{fig:1d_ska}
\end{figure*}

The final metric of interest for 21~cm studies, where the cosmological signal is expected to be isotropic, is the spherically-averaged (1D) power spectrum. To remove the bulk of the foreground extended source bias, we consider line-of-sight scales larger than $k_\parallel=0.1$ (smaller spatial scales). In doing so, foregrounds are reduced when averaging spherically, but we also lose spatial modes. Figure \ref{fig:1d_ska} displays these 1D profiles, obtained directly from the power on each measured baseline at each frequency (i.e., not obtained from 2D, but considering the original 3D distribution).
The same conclusions can be drawn about the relative merits of each array at the lowest frequency (50~MHz). The different structural properties stem from the interaction of the Blackman-Nuttall spectral taper and each array's spectral covariance profiles. Fundamentally, it is the progression in $k$-mode of the additional leaked power as the arrays become more compact, that is of concern. Here we also provide results from the 39~km maximum baseline, but using the v7 proposed array of \citet{jones16} with unpacked clusters (green dashed). Here, the addition of $uv$ information on intermediate scales provides some improvement relative to the Baseline Design, and can recover information lost from cutting from 49~km to 39~km maximum baselines.

It is important to understand where the differences in the arrays are contributing to the differences in the performance. Intuitively, sources that are extended, but have power on angular scales smaller than that for the EoR, will not contribute power bias into the EoR parameter space. Similarly, sources that have power on large angular scales, but no power on small scales, contribute to EoR baselines, but each of the 39~km, 49~km and 65~km arrays sample these scales equally. Therefore, we hypothesise that it is sources that have both large and small-scale structures that couple inability to measure small scales into the estimation of the larger scales. To test this hypothesis, we perform the same analysis at 50~MHz with two additional sources: one source with power only above 1~arcmin scales (broad source), and one source with power only on scales of 10~arcsec--36~arcsec (compact source). When comparing the 39~km and 65~km arrays for their power bias in the EoR power spectrum, we find that both yield comparable performance (power ratio approximately unity). This supports the hypothesis that sources that have multi-scale power are those that show differences in estimation precision for the different arrays.

\subsection{Discussion and interpretation}
The key message of this work is that a sky model generated from a given array, and subsequently used as `truth' (the reference model for a source for calibration and subtraction), has inherent errors, due limitations of that array, and that those errors will propagate forward into science data products. Here, we considered a generalised multi-scale Gaussian source (extended source), and used mock SKA1-Low and existing and upcoming MWA+TGSS array configurations to estimate their model parameters, in the presence of a realistic sky with structured sidelobes from other sky sources. We applied these errors to model EoR data, and quantified the bias in power in the EoR 2D and 1D power spectra due to these incomplete source models. Compared with previous work in this area, the use of a generalised extended source model, the inclusion of other confusing sources in the field (classical and sidelobe confusion), and the direct equating of modes measured by a survey to input calibration model `truth' parameters, expands on existing studies.

For the MWA EoR experiment, we find that both the additional short baseline $uv$-coverage, and the additional MWA256+TGSS long baselines, are required to obtain sufficient information on all relevant scales, for a high signal-to-noise ratio detection.

For SKA1-Low, we find that arrays with maximum baselines of 49~km and 65~km yield comparable performance at 50~MHz and 150~MHz, while 39~km, 14~km and 4~km arrays yield degraded performance. This is particularly true at Cosmic Dawn (low) frequencies, where SKA1 is aiming to be transformational and new. We therefore conclude that 49~km maximum baselines are sufficient to form the sky and calibration model for EoR/CD science, \textit{but 39~km baselines, corresponding to removal of two clusters from each spiral arm, yield degraded results and threaten high-redshift Cosmic Dawn science}. This is not true for the same longest baseline but an array with more intermediate scales (i.e., the v7 design). Such alterations of the inner array may be used to alleviate some of the degradation caused by removing the outer stations. We additionally find that it is multi-scale sources, which have power on large and small scales, that are those that yield different power bias for the 39~km and 65~km arrays. It is these sources for which the sky model estimation will yield differences depending on array design.

\section{Summary and forward look}
The broad design of SKA1-Low is relatively fixed, with an expectation of an aperture array interferometer of $\sim$130,000 dipoles collected into $\sim$500 stations, spanning tens of kilometres, and with a densely-packed core of 50--60\% of the collecting area within 2--3~km \citep{dewdney16}. This broad model provides the exceptional surface brightness sensitivity and wide frequency coverage to address the exciting science goals of the observatory. The specific details of station location and maximum array baseline are under discussion. To contribute to that discussion, here we study the impact of incomplete sky models of extended sources on EoR and CD science for MWA and SKA. The recommendations for the design of SKA1-Low are (1) an unpacking of inner station clusters; (2) a minimum longest baseline of 50~km. Combining these recommendations may improve performance but at an overall reduced cost.

\begin{acknowledgements}
CMT thanks Robert Braun and Ben McKinley for useful discussions.
The Centre for All-Sky Astrophysics (CAASTRO) is an Australian Research Council Centre of Excellence, funded by grant CE110001020. The Centre for All-Sky Astrophysics in 3D (ASTRO 3D) is an Australian Research Council Centre of Excellence, funded by grant CE170100013.  This research has made use of NASA's Astrophysics Data System. CMT is supported under the Australian Research Council's Discovery Early Career Researcher funding scheme (project number DE140100316). This work was supported by resources provided by the Pawsey Supercomputing Centre with funding from the Australian Government and the Government of Western Australia. We acknowledge the International Centre for Radio Astronomy Research (ICRAR), a Joint Venture of Curtin University and The University of Western Australia, funded by the Western Australian State government.
\end{acknowledgements}

\bibliographystyle{pasa-mnras}
\bibliography{references}

\end{document}